\documentclass[runningheads,orivec]{llncs}
\usepackage{graphicx}
\usepackage{hyperref}
\usepackage{amssymb}
\usepackage{amsmath}
\usepackage{scrextend}
\usepackage[linesnumbered,ruled]{algorithm2e}
\usepackage{tabularx}
\usepackage{multirow}
\usepackage{mathtools}
\usepackage{color}
\usepackage{booktabs}
\usepackage[font=small,labelfont=bf]{caption}
\usepackage{pifont}

\usepackage[overload]{empheq}
\usepackage{threeparttable}
\usepackage{framed}

\usepackage{xspace}
\newcommand{\etal}{\textit{et al.}\xspace}
\newcommand{\ie}{\textit{i.e.,}\xspace}
\newcommand{\eg}{\textit{e.g.,}\xspace}

\hypersetup{
	colorlinks,
	citecolor= blue,
	urlcolor= blue
}

\newcommand\ttt[1]{\texttt{#1}}
\newcommand\tsf[1]{\textsf{#1}}

\SetKw{KwPublic}{$public$ Input:}
\SetKw{KwPrivate}{$private$ Input:}
\SetKw{KwOutput}{Output:}
\SetKwFunction{KwFn}{Function}
\SetKw{KwStat}{Statement $\vec{s}$:}
\SetKw{KwWit}{Witness $\vec{w}$:}

\let\oldnl\nl
\newcommand{\nonl}{\renewcommand{\nl}{\let\nl\oldnl}}

\usepackage{xcolor}

\newcommand{\mg}[1]{\textcolor{black}{#1}}
\usepackage[normalem]{ulem}

\begin{document}

\title{Dispute-free Scalable Open Vote Network using zk-SNARKs}

\author{Muhammad ElSheikh\inst{1,2} \and Amr M. Youssef\inst{1}}
\authorrunning{M. ElSheikh et al.}
\institute{Concordia Institute for Information Systems Engineering, \\Concordia University, Montr\'{e}al, Qu\'{e}bec, Canada\\
\and
National Institute of Standards (NIS), Cairo, Egypt\\
	\email {\{m\_elshei,youssef\}@ciise.concordia.ca}
}
\maketitle



\begin{abstract}
	The Open Vote Network is a self-tallying decentralized e-voting protocol suitable for boardroom elections. Currently, it has two Ethereum-based implementations: the first, by McCorry \etal,  has a scalability issue since all the computations are performed on-chain. The second implementation, by Seifelnasr \etal,  solves this issue partially by assigning a part of the heavy computations to an off-chain untrusted administrator in a verifiable manner. As a side effect, this second implementation became not dispute-free; there is a need for a tally dispute phase where an observer interrupts the protocol when the administrator cheats, \ie announces a wrong tally result. In this work, we propose a new smart contract design to tackle the problems in the previous implementations by (i) preforming all the heavy computations off-chain hence achieving higher scalability, and (ii) utilizing zero-knowledge Succinct Non-interactive Argument of Knowledge (zk-SNARK) to verify the correctness of the off-chain computations, hence maintaining the dispute-free property. 
	\mg{To demonstrate the effectiveness of our design, we develop prototype implementations on Ethereum and conduct multiple experiments for different implementation options that show a trade-off between the zk-SNARK proof generation time and the smart contract gas cost, including an implementation in which the smart contract consumes a constant amount of gas independent of the number of voters.}
\end{abstract}


%
%
\section{Introduction}
E-voting refers to an election system in which voters can cast their vote electronically.  The main advantages of e-voting, compared to the traditional paper-based election, include high speed of tallying, cost-effectiveness, and scalability. Using e-voting systems can be crucial in many situations, \eg the current  COVID-19 pandemic renders traditional paper-based voting within organizations a potential health hazard and sometimes not possible because of the work from home setup.
Nowadays, there are many e-voting systems that can support a number of voters from a boardroom to a national scale \cite{GRITZALIS2002539,Maaten2004,adida2008helios}. However, 
most of them rely heavily on a trusted central authority, which might lead to violating the voters' privacy. With the emerging of blockchain technology as a decentralized append-only ledger, many researchers have proposed several blockchain-based e-voting protocols (\eg, see \cite{8405627,9149825,9031381,mccorry,murtaza2019blockchain,seifelnasr}). 
Unfortunately, widely deployed blockchains such as Bitcoin and Ethereum suffer from scalability issues. Moreover, they do not inherently provide the privacy required by e-voting protocols. Therefore, a good blockchain-based e-voting system should handle these limitations.  

The Open Vote Network is a self-tallying decentralized voting protocol. Self-tallying  means that anyone who observes the protocol can tally the result without counting on a trusted authority. The protocol also provides maximum voter privacy; a single vote can only be breached by a full-collusion involving compromising all other votes.
McCorry \etal \cite{mccorry} presented the first implementation of the Open Vote Network protocol on the Ethereum blockchain. However, their implementation does not provide  scalability because all the protocol computations are delegated to the smart contract. This problem is partially solved by Seifelnasr \etal \cite{seifelnasr} by assigning the tallying computations to an off-chain untrusted administrator in a verifiable manner. To address the possibility of a malicious administrator, they added a dispute phase in which  an honest voter may interrupt the protocol if the administrator \mg{provides an incorrect} tallying result. As a result, the protocol lost its dispute-free property.

\subsubsection{Contribution.} In this work, we provide a new design to deploy the Open Vote Network using Ethereum smart contract. The new design can achieve better scalability without loosing the dispute-free property. Our contribution can be summarized as follows.
\begin{enumerate}
	\item We develop a smart contract for the Open Vote Network in which all the heavy computations are performed off-chain without loosing dispute-free property.
	\item We design three zk-SNARK arithmetic circuits to verify that all the off-chain computations are performed correctly by their responsible parties.
	\item We develop a prototype\footnote{\url{https://github.com/mhgharieb/zkSNARK-Open-Vote-Network}} of our design to assess its performance. We also conduct some experiments to estimate the maximum number of voters that can be supported before exceeding the gas limit of the Ethereum block.
	\mg{\item Finally, we show how to enhance the scalability of our design by modifying the zk-SNARK circuits such that they have statements of a fixed size. Consequently, the smart contract functions which verify the correctness of these zk-SNARK proofs consume fixed gas cost independent of the number of voters. The tradeoff between the zk-SNARK proof generation time and the smart contract gas cost is also experimentally evaluated.
	} 
\end{enumerate}

\par The rest of the paper is organized as follows. In Section \ref{sec:related}, we briefly  revisit some related work on voting protocols implemented on the Ethereum blockchain. Section \ref{sec:preliminaries} recalls the cryptographic primitives utilized in our protocol. In Section \ref{sec:design}, we  provide our design of the zk-SNARK circuits and the smart contracts. In Section \ref{sec:evaluation}, we evaluate our design and compare it against previous work. \mg{In Section \ref{furtherImprovement}, we provide multiple enhancements to the design in order to achieve better scalability, with different trade-offs between proof generation time and gas cost. }Finally, our conclusion is presented in Section \ref{sec:conclusion}.  
\section{Related Work}\label{sec:related}
The Open Vote Network is a self-tallying decentralized e-voting protocol. The concept of self-tallying was introduced by Kiayias and Yung \cite{10.1007/3-540-45664-3_10} for boardroom voting. This work was followed by Groth \etal \cite{groth2004efficient} and Hao \etal \cite{hao2010anonymous} who proposed a system that provides better efficiency for each voter. Hao \etal's protocol has the same security properties and achieves better efficiency in terms of number of rounds. Li \etal \cite{9031381} presented a new blockchain based self-tallying voting protocol for decentralized IoT. Recently, Li \etal \cite{9149825} proposed a self-tallying protocol that utilizes homomorphic time-lock puzzles to encrypt the votes for a specified duration of time to maintain the privacy of ballots during the casting phase. 

McCorry \etal \cite{mccorry}  and Seifelnasr \etal \cite{seifelnasr} presented two implementations \mg{of the system} of Hao \etal \cite{hao2010anonymous} as smart contracts on Ethereum. As mentioned above, the first implementation  suffers from a scalability issue and the second requires a third-party to observe the behavior of the election administrator.

\section{Preliminaries}\label{sec:preliminaries}
\subsection{zk-SNARK}
A zk-SNARK refers to a zero-knowledge Succinct Non-interactive Argument of Knowledge scheme which enables a prover to convince a verifier that a statement is true without prior interactions between them \cite{groth2010short}. 

Suppose an arithmetic circuit $C$ with a relation $\mathcal{R}_C$ and a language $\mathcal{L}_C$ takes as input a statement $\vec{s}$ and a witness $\vec{w}$ s.t. $(\vec{s},\; \vec{w}) \in \mathcal{R}_C$. A zk-SNARK for this arithmetic circuit satisfiability is defined by the following triple of polynomial-time algorithms \cite{groth2010short,groth16,parno2013pinocchio}: 

\begin{itemize}
	\item (\tsf{pk}, \tsf{vk}) $\leftarrow \tsf{Setup}(1^\lambda,\;C)$. Given a security parameter $\lambda$ and the circuit $C$, the algorithm generates a common reference string \mg{(CRS)} that contains a pair of keys; a proving key \tsf{pk}  and a verifying key \tsf{vk}. Both keys are considered as public parameters for the circuit $C$. 
	\item $\pi \leftarrow \tsf{Prove}(\tsf{pk},\;\vec{s},\; \vec{w})$.  Given a proving key \tsf{pk}, a statement $\vec{s}$, and a witness $\vec{w}$ s.t. $(\vec{s},\; \vec{w}) \in \mathcal{R}_C$, the algorithm generates a zero-knowledge non-interactive proof $\pi$ for the statement $\vec{s} \in \mathcal{L}_C$ that reflects the relation between $\vec{s}$ and $\vec{w}$.
	\item $0/1 \leftarrow \tsf{Verify}(\tsf{vk},\; \vec{s},\; \pi)$. Given a verifying key \tsf{vk}, the statement $\vec{s}$, and the proof $\pi$, the algorithm outputs 1 if $\pi$ is a valid proof for the statement $\vec{s} \in \mathcal{L}_C$, and outputs 0 otherwise.
\end{itemize} 

\noindent Typically, a zk-SNARK provides the following security properties \cite{groth16}:
\begin{enumerate}
	\item \textbf{Perfect Completeness}: For each valid statement $\vec{s}$ with a valid witness $\vec{w}$ s.t. $(\vec{s},\; \vec{w}) \in \mathcal{R}_C$, an honest prover always convinces an honest verifier, \ie $\tsf{Verify}(\tsf{vk},\; \vec{s},\; \pi)$ outputs 1 with a probability equal to 1.
	\item \textbf{Computational Soundness}: A polynomial-time malicious prover cannot convince the verifier of a false statement, \ie $\tsf{Verify}(\tsf{vk},\; \vec{s},\; \pi)$ outputs 1 with a probability $\approx 0$ when the statement $\vec{s} \notin \mathcal{L}_C$.
	\item \textbf{Computational Zero-Knowledge}: A polynomial-time adversary cannot extract any information about the witness from the honestly-generated proof.
	
	\item \textbf{Succinctness}. A zk-SNARK is succinct if the honestly-generated proof size is polynomial in $\lambda$ and $\tsf{Verify}(\tsf{vk},\; \vec{s},\; \pi)$ runs in polynomial time in $\lambda + |\vec{s}|$.
\end{enumerate}

\subsection{Open Vote Network}
The Open Vote Network is a decentralized two-round self-tallying e-voting protocol \cite{hao2010anonymous}. It is suitable for a boardroom election in which the number of voters is relatively small.

In the beginning, eligible voters ($\mathcal{P}_0, \mathcal{P}_1, \ldots, \mathcal{P}_{n-1}$) agree on a finite cyclic group $\mathbb{G}$ of a prime order $q$ and a generator $g$ in which the Decisional Diffie-Hellman (DDH) problem is intractable. Then, each voter $\mathcal{P}_i$ picks a random value $x_i \in_R \mathbb{Z}_q$ as her private voting key. The Open Vote Network is executed for an election with two options 1 or 0 (implying `YES' or `NO') as follows.

\textbf{Round 1.} Each eligible voter $\mathcal{P}_i$ publishes her public voting key $g^{x_i}$ along with a non-interactive zero-knowledge proof of knowledge regarding her private voting key $x_i$ on the public bulletin board. At the end of this round, each voter verifies the validity of other voters' zero-knowledge proof of knowledge, then computes her blinding key $Y_i$ as in Eq. \ref{eq1}. 

\begin{equation}\label{eq1}
	Y_{i}=\prod_{j=0}^{i-1}g^{x_{j}}/\prod_{j=i+1}^{n-1}g^{x_{j}}
\end{equation}

By implicitly setting $Y_{i} = g^{y_{i}}$, it is easy to prove that $\prod_{i}Y_i^{x_i} = g^{\sum_{i}x_iy_i} = g^0=1$.

\textbf{Round 2.} Each eligible voter $\mathcal{P}_i$  uses her blinding key $Y_i$ and the private key $x_i$ to encrypt the vote $v_i \in \{0,1\}$ s.t. the encrypted vote $V_i = g^{v_i} Y_i^{x_i}$. Then she publishes the encrypted vote $V_i$ along with a non-interactive zero-knowledge proof of validity to prove that the encrypted vote $V_i$ is well-formed such that $v_i \in \{0,1\}$. At the end of this round and after verifying the non-interactive zero-knowledge proofs of all encrypted votes, anyone who observes the protocol can compute the tally of `YES' votes by exploiting the homomorphic property in the encrypted votes as follows: 
	$\prod_i V_i =\prod_{i}g^{x_{i}y_{i}}g^{v_{i}}= g^{\sum_i  x_iy_i + v_i}=g^{\sum_i v_i}$.
Accordingly, the tally result of `YES' votes $\sum_i v_i$ can be easily obtained by performing an exhaustive search on the discrete log of $g^{\sum_i v_i}$. This exhaustive search is bounded by the number of voters which is relatively small. For more details, see \cite{hao2010anonymous}.

\section{Protocol Design}\label{sec:design}
In this section, we present our proposed design to deploy the Open Vote Network on the Ethereum blockchain using zk-SNARKs.

\subsection{zk-SNARK Arithmetic Circuit}
\mg{Since validating zk-SNARK proofs on Ethereum are performed over two cyclic groups of prime order $p$ \cite{eip-197}, our protocol computations are preformed over a cyclic group $\mathbb{G}$ with a finite field $\mathbb{F}_p$ on the elliptic curve \textit{Baby Jubjub} \cite{math9233022}. $\mathbb{G}$ has a prime order $q$,} a base point (generator) $G$, and a point at infinity (the neutral element) $O$. A point $P \in \mathbb{G}$ is presented by its two coordinate values $(P^x, P^y)$.

In our design, we use zk-SNARKs to verify that all off-chain computations are performed correctly by their responsible parties. To this end, we design three zk-SNARK arithmetic circuits: \tsf{publicKeyGen}, \tsf{encryptedVoteGen}, and \tsf{Tallying} corresponding to generating the public key of voters, encrypting the votes, and tallying the result of `YES' option, respectively. 

We design these circuits based on Groth16 zk-SNARK construction \cite{groth16} because it is a quadratic arithmetic program (QAP) hence it provides a linear-time \tsf{Setup}, quasilinear-time \tsf{Prove}, and linear-time \tsf{Verify} \cite{sasson2014zerocash}. However, Groth16 enforces some restrictions on the design (\eg array indices and loop iteration counts must be constant during compiling (\tsf{Setup}) phase).

During the design, we use the following pre-defined arithmetic circuits as building blocks:
\begin{itemize}
	\item \tsf{Mux($s, P, Q$)}: Returns $P$ if the selector $s=0$, and $Q$ if $s=1$.
	\item \tsf{LessThan($a,b$)}: Returns 1 if $a<b$, and 0 otherwise.
	\item \tsf{GreaterThan($a,b$)}: Returns 1 if $a>b$, and 0 otherwise.
	\item \mg{\tsf{CompC($a,c$)} where $c$ is a constant: Returns 1 if $a>c$, and 0 otherwise.}
	\item \mg{\tsf{Bits2Num($a_0,\ldots,a_{k-1}$)}: Returns the integer number represented by bits $a_0,\ldots,a_{k-1}$.}
	\item \tsf{IsPoint($x,y$)}: Returns 1 if the pair ($x,y$) is a point on the elliptic curve, and 0 otherwise.
	\item \tsf{IsEqual($ P $, $ Q $)}: Returns 1 if the two points $P$ and $Q$ are equal, and 0 otherwise. 
	\item \tsf{eADD($P, Q$)}: Point addition ($P + Q$) on the elliptic curve.
	\item \tsf{eSUB($P, Q$)}: point subtraction ($P - Q$) on the elliptic curve.
	\item \tsf{eScalarMUL($a$,$P$)}: Scalar multiplication ($aP$) on the elliptic curve.
\end{itemize}

\mg{
Let $\kappa =|p| - 1$. Since all operations are preformed over $\mathbb{F}_p$, the number of inputs to \tsf{Bits2Num} must be  $\le \kappa$ bits to avoid overflow when calculating the output value.}

\subsubsection{\tsf{publicKeyGen} Circuit ($C_{PK}$).} \mg{This circuit (see Circuit \ref{alg:CPK}) is intended for the setup of the zk-SNARK to prove that}  a voter \ttt{B} knows the private key $x_B$ corresponding to the public key $PK_{B} = x_B G$ \mg{ in step (1); and  the sign of its $x$-coordinate is 0 (\ie $pk^{x_B}_B < p/2$), in step (3), to ensure that $PK_{B}$ follows the compact representation as described in \cite{jivsov_compact_2014} such that $x$ and $y$ coordinates have a one-to-one relation}. 

\subsubsection{\tsf{encryptedVoteGen} Circuit ($C_{V}$).} \mg{This circuit is intended for the setup of the zk-SNARK to prove that} a voter \ttt{B} with index $i_B$ forms her encrypted vote $V_B$ correctly s.t. the vote $v_B \in \{0,1\}$ as shown in Circuit \ref{alg:CV}. 

Since the verification time is linearly proportional to the size of the statement $\vec{s}$, we decompose each public key ${PK}_i$ into its coordinate values ($pk_i^x, pk_i^y$). \mg{Since they have a one-to-one relation,} the $y$-coordinate becomes a part of the statement $\vec{s}$ and the $x$-coordinate becomes a part of the witness $\vec{w}$
in order to optimize the circuit and reduce the verification time, hence the on-chain computation.

From Eq. \ref{eq1}, the encrypted vote $V_B$ is computed as follows:
\begin{align}
	V_B &= v_B G + x_B Y_B \label{eq2}\\
	Y_B &= \underbrace{\sum_{i=0}^{i_B -1} PK_i}_{Y_l} - \underbrace{\sum_{i=i_B + 1}^{n-1} PK_i}_{Y_g} \label{eq3}
\end{align}

Accordingly, \tsf{encryptedVoteGen} circuit checks that
$v_B \in \{0,1\}$ in step (1); each pair ($pk_i^x, pk_i^y$) is a point on the elliptic curve in step (5); \mg{ the sign of the $x$-coordinate is 0 ($pk^{x_B}_B < p/2$) in step (6) to verify the one-to-one relation;} the correctness of $Y_l$, $Y_g$, and $Y_B$ (as in Eq. \ref{eq3}) in steps (8-10), (11-13), and (15), receptively; and the correctness of $V_B$ (as in Eq. \ref{eq2}) in steps (16-18).

\subsubsection{\tsf{Tallying} Circuit ($C_{T}$).} \mg{This circuit is intended for the setup of the zk-SNARK to prove} the correctness of  the tallying result as shown in Circuit \ref{alg:CT}. Similar to $C_V$, each encrypted vote ${V}_i$ is decomposed into its coordinate values ($V_i^x, V_i^y$), then the $y$-coordinate becomes a part of the statement $\vec{s}$ and the $x$-coordinate becomes a part of the witness $\vec{w}$
in order to reduce the on-chain computation. \mg{Since we cannot enforce the compact representation for $V_i$, we instead present the sign of $x$-coordinate in a single bit $S_i$ then compress every $\kappa$ sign-bits in one integer number $D_j$. Hence, we can verify the one-to-one relation between the two coordinates of $V_i$.}

Accordingly, \tsf{Tallying} circuit checks that each pair ($V_i^x, V_i^y$) is a point on the elliptic curve in step (6)\mg{. Then, it checks the correctness of } $\sum_{i}V_i$ in step (7); \mg{the $x$-coordinate sign ($S_i$) of $V_i$ in step (9); the compression of every $\kappa$ sign-bits into an integer $D_j$ in step (12);} the incrementally exhaustive search in steps (17-19); and finally the tallying result \mg{($res$) s.t. $\sum V_i = (res)G$ in steps (21-22). It should be mentioned that $0 \le res \le n$, where $res = 0 $ when no voter selects the `Yes' option and $res = n$ when all voters select the `Yes' option. Therefore, the exhaustive search counter $i$ in step (16) starts from 0 to $n$.}

\noindent
\begin{minipage}{.5\textwidth}
\begin{algorithm}[H]
	\DontPrintSemicolon
	\caption{\tsf{publicKeyGen}}
	\label{alg:CPK}
	\nonl\KwStat{$PK_B$}\;
	\nonl\KwWit{$x_B$}\;
	$PK_B \gets \textsf{eScalarMUL}(x_B,G)$\;
	\mg{$({pk}_B^{x}, {pk}_B^{y})\leftarrow PK_B$\;}
	\mg{\ttt{Assert} \tsf{CompC$({pk}_B^{x}, p/2) = 0$}\;}
	
\end{algorithm}
	\begin{algorithm}[H]
		\DontPrintSemicolon
		\SetArgSty{textnormal}
		\caption{\tsf{encryptedVoteGen}}
		\label{alg:CV}
		\nonl\KwStat{($V_B$, $i_B$, \mg{$\{{pk}_i^{y}\}$})}\;
		\nonl\KwWit{($v_B$, $x_{B}$, \mg{\{${pk}_i^{x}$\}})}\;
		\ttt{Assert $(1-v_B) \times v_B = 0$}\;
		$Y_l \gets O$\;
		$Y_g \gets O$\;
		
		\For{$i \gets 0$ \KwTo $ n-1$}{
			\ttt{Assert} \tsf{IsPoint}(${pk}_i^x,\;{pk}_i^y$)\;
			\mg{\ttt{Assert} \tsf{CompC$({pk}_i^{x}, p/2) = 0$}}\; 
			$PK_i \leftarrow ({pk}_i^x,\; {pk}_i^y)$\;
			$e_l \gets$ \textsf{LessThan($i$, $i_B$)}\;
			$T_l \gets \tsf{Mux}(e_l,\;O,\;PK_i)$\;
			$Y_l \gets \textsf{eADD}(Y_l, T_l)$\;
			$e_g \gets$ \textsf{GreaterThan($i$, $i_B$)}\;
			$T_g \gets \tsf{Mux}(e_g,\;O,\;PK_i)$\;
			$Y_g \gets \textsf{eADD}(Y_g, T_g)$
		}
		$Y_B \gets \textsf{eSUB}(Y_l, Y_g)$\;
		$T_0 \gets \textsf{eScalarMUL}(x_B, Y_B)$\;
		$T_1 \gets \tsf{Mux}(v_B,\;O,\;G)$\;
		$V_B \gets \textsf{eADD}(T_0, T_1)$	
	\end{algorithm}
\end{minipage}
\begin{minipage}{.5\textwidth}
	\begin{algorithm}[H]
		\DontPrintSemicolon
		\SetArgSty{textnormal}
		\caption{\tsf{Tallying}}
		\label{alg:CT}
		\nonl\KwStat{($res$,\mg{\{$D_j$\}},\mg{\{$V_i^y$\}})}\;
		\nonl\KwWit{\mg{\{$V_i^x$\}}}\;
		
        \mg{$l \gets \lceil \frac{n}{\kappa} \rceil$\;}
        \mg{$\{S_i \gets 0, 0 \leq i \leq \kappa l-1\}$\;}
        \mg{$\{D_j, 0 \leq j \leq l-1\}$\;}
        $sumV \gets O$\;
		\For{$i \gets 0 $ \KwTo $ n-1$}{
			\ttt{Assert} \tsf{IsPoint}(${V}_i^x,\;{V}_i^y$)\;
			$sumV \gets \textsf{eADD}(sumV, V_i)$\;
			$V_i \leftarrow ({V}_i^x,\; {V}_i^y)$\;
			\mg{$S_i \gets \tsf{CompC}({V}_i^{x}, p/2)$}\;
			}
		\mg{\For{$j \gets 0$ \KwTo $l-1$}{
		    $D_j \gets \tsf{Bits2Num}(S_{\kappa j},\ldots, S_{\kappa j+\kappa -1})$
		}}
		$t \gets 0$\;
		$T \gets O$\;
		\For{$i \gets 0 $ \KwTo $ n$}{
			$e \gets \textsf{IsEqual}(T, sumV)$\;
			$t \gets t + e \times i$\;
			$T \gets \textsf{eADD}(T, G)$\;
		}
		\ttt{Assert} $\textsf{eScalarMUL}(t,G) = sumV$\;
		$res \gets t$\; 
				\nonl\;
	\end{algorithm}
\end{minipage}

\subsection{Open Vote Network Smart Contract}
Before executing the protocol, an administrator \ttt{A} and a set of $n$ eligible voters run an MPC-based setup ceremony for generating the proving and verifying keys for the arithmetic circuits.
\begin{align*}
	(\tsf{pk}_{C_{PK}}, \tsf{vk}_{C_{PK}}) &\leftarrow \tsf{Setup}(1^\lambda, C_{PK})\\
	(\tsf{pk}_{C_{V}}, \tsf{vk}_{C_{V}}) &\leftarrow \tsf{Setup}(1^\lambda, C_{V})\\
	(\tsf{pk}_{C_{T}}, \tsf{vk}_{C_{T}}) &\leftarrow \tsf{Setup}(1^\lambda, C_{T})\\
\end{align*}

After that, similar to \cite{seifelnasr}, the administrator accumulates the list of eligible voters in a Merkle tree $MT_{\mathcal{E}}$  \mg{where the voters' Ethereum account addresses are the tree leaves,} and publishes it on IPFS 
\mg{allowing each voter to generate her proof of voting eligibility}. Also, the administrator defines the time intervals of the protocol phases, namely, Registering Voters, Casting Encrypted Votes, Tallying the Result, and Refunding.  

\subsubsection{Smart Contract Deployment.}
Subsequently, the administrator deploys the smart contract, which initializes the variables used in the subsequent phases,  with the following set of parameters \mg{and pays a collateral deposit $F$} as shown in Fig. \ref{figDeploy}.
\begin{figure}[htbp]
	\begin{framed}
		\centering
		\begin{tabular}{p{.2\linewidth}p{.8\linewidth}}
			\tsf{Deploy:}& upon receiving $(root_{\mathcal{E}},\; \tsf{vk}_{C_{PK}},\; \tsf{vk}_{C_{V}},\; \tsf{vk}_{C_{T}},\; T_{1},\; T_{2},\; T_{3},\;T_{4},\; n)$ from administrator \texttt{A}:\\ 
			&   \texttt{Assert $value = F$}\\
			& \texttt{Set $admin:= \texttt{A}$}\\
			&   \texttt{Store $root_{\mathcal{E}}, \tsf{vk}_{C_{PK}}, \tsf{vk}_{C_{V}}, \tsf{vk}_{C_{T}}, T_1, T_{2}, T_{3},T_{4}, n$}\\
			&   \texttt{Init} $voters:= \{\}$, $publicKeys:= \{\}$, $EncryptedVotes := \{\}$,  \mg{$Int\_Vsigns:= \{0,\ldots,0\}$,} $tallyingResult:=\ttt{NULL}$, $index:=0$
		\end{tabular}
	\end{framed}
	\caption{Pseudocode for deployment of the smart contract.}
	\label{figDeploy}
\end{figure}

\begin{itemize}
	\item $root_{\mathcal{E}}$: The root of the Merkle tree $MT_{\mathcal{E}}$.
	\item $\tsf{vk}_{C_{PK}}, \tsf{vk}_{C_{V}}, \tsf{vk}_{C_{T}}$: The verifying keys of the zk-SNARK circuits.  
	\item $T_{1}, T_{2}, T_{3}, T_{4}$: The block heights that define the end of the protocol phases.
	\item $n$: The number of eligible voters.
	\item \textit{$F$}: A collateral deposit paid by both the administrator and the voters to penalize malicious actors.
	\mg{\item $voters,\; publicKeys,\; EncryptedVotes,\;Int\_Vsigns$: Four arrays of sizes $n,\;n,\;n,$ and $l=\lceil\frac{n}{\kappa}\rceil$, used to store the voters' Ethereum account addresses, $\{PK_i\}$, $\{V_i\}$, and $\{D_j\}$, respectively.}
	
\end{itemize}

\subsubsection{Registering Voters.} This phase starts immediately after deploying the contract. Each voter \ttt{B} \mg{selects} her private key $x_B$ and uses \tsf{publicKeyGen} circuit along with $\tsf{pk}_{C_{PK}}$ to generate the public key $PK_B$ and a zk-SNARK proof $\pi_{x_B}$.
After that, she invokes \tsf{Register} function in the smart contract with the parameters $PK_B$ and $\pi_{x_B}$ along with a Merkle proof of voters membership $\pi_B$\mg{, and pays the collateral deposit $F$} as shown in Fig. \ref{figRegister}. On its turn, the function ensures that the voter deposits the correct collateral fee, it is invoked within the allowed interval, and the number of already registered voters does not exceed the total number of eligible voters. After that, it reconstructs the zk-SNARK statement $\vec{s}$ of \tsf{publicKeyGen} circuit. Consequently, \tsf{Register} verifies both the Merkle tree proof of the voter membership and the zk-SNARK proof using $\tsf{vk}_{C_{PK}}$ key. We utilize the implementation \mg{ in \cite{galal2019efficient} to verify} the Merkle proof. Finally, it stores the public key $PK_B$ and the address of the voter \ttt{B} for the subsequent phases.
\begin{figure}[htbp]
	\begin{framed}
		\centering
		\begin{tabular}{p{.2\linewidth}p{.8\linewidth}}
			\tsf{Register:} & upon receiving $(PK_{B}, \pi_{x_B}, \pi_{B})$ from voter \texttt{B}:  \\ 
			&   \texttt{Assert $value = F$}\\
			&   \texttt{Assert $T < T_{1}$}\\
			& \ttt{Assert $index < n$}\\
			&   \texttt{Reconstruct $\vec{s}: = PK_{B}$}\\
			&   \texttt{Assert MerkleTree.verify$(\;\pi_{B},\; \texttt{B},\; root_{\mathcal{E}})$}\\
			&   \texttt{Assert zkSNARK.verify($\tsf{vk}_{C_{PK}}$, $\vec{s}$, $\pi_{x_B}$)}\\
			&   \texttt{Store $publicKeys[index] := PK_{B}$}\\
			&   \texttt{Store $voters[index] := \ttt{B}$}\\
			& \texttt{Set $index := index +1$}\\
		\end{tabular}
	\end{framed}
	\caption{Pseudocode for \tsf{Register} function}
	\label{figRegister}
\end{figure}

\subsubsection{Casting Encrypted Votes.}
After the voters registration phase ends, each voter \ttt{B}, which has index $i_B$, starts computing her blinding key $Y_B$ to encrypt her vote $v_B$ and generate a zk-SNARK proof $\pi_{v_B}$ using \tsf{encryptVoteGen} circuit along with the proving key $\tsf{pk}_{C_V}$. Then, she publishes the encrypted vote $V_B$ publicly by invoking \tsf{CastVote} function in the smart contract. The function verifies that the voter casts the encrypted vote within the allowed time interval, and the voter indeed has the index $i_B$. After that, it reconstructs the statement $\vec{s}$ of \tsf{encryptedVoteGen} circuit along with $\tsf{vk}_{C_V}$ in order to verify the correctness of the zk-SNARK proof $\pi_{v_B}$ submitted by the voter. Finally, it stores the encrypted vote $V_B$ \mg{and updates $Int\_Vsigns$ based on the sign of its $x$-coordinate, $V_B^x$, and the voter index $i_B$} as depicted in Fig. \ref{figCast}.

\begin{figure}[!ht]
	\begin{framed}
		\centering
		\begin{tabular}{p{.2\linewidth}p{.8\linewidth}}
			\tsf{CastVote:}& upon receiving $(V_B,\; i_B,\; \pi_{v_B} )$ from voter \texttt{B}\\ 
			&   \texttt{Assert $T_{1}$ < $T$ < $T_{2}$}\\
			& \texttt{Assert B = $voters[i_B]$}\\
			&   \texttt{Reconstruct} $\vec{s}: = (V_B,\;  i_B,\; publicKeys)$\\
			&   \texttt{Assert zkSNARK.verify}($\tsf{vk}_{C_V}$, $\vec{s}$, $\pi_{v_{B}}$)\\
			&   \texttt{Set $EncryptedVotes[i_B] :=V_B$}\\
		    & \mg{\ttt{IF}($V_B^x > p/2$):}\\ 
		&\mg{~~~~~~~\texttt{Set} $Int\_Vsigns[\lfloor \frac{i_B}{\kappa}\rfloor]  := Int\_Vsigns[\lfloor \frac{i_B}{\kappa}\rfloor] \oplus 2^{i_B \mbox{ mod } \kappa} $}
		\end{tabular}
	\end{framed}
	\caption{Pseudocode for \tsf{CastVote} function}
	\label{figCast}
\end{figure}

\subsubsection{Tallying the Result.} After the casting phase ends, the administrator obtains all the encrypted votes stored in the smart contract in order to tally the result of `YES' option. To this end, she uses \tsf{Tallying} circuit along with $pk_{C_T}$ to obtain the result $res$ and its corresponding zk-SNARK proof $\pi_{res}$. Then, she invokes \tsf{SetTally} function to publish this result. The function verifies that the transaction is within the allowed time interval and it is sent by the administrator who deployed the smart contract. Then, the function reconstucts the statement $\vec{s}$ of \tsf{Tallying} circuit to verify the correctness of the tallying result as shown in Fig. \ref{figTally}.

\begin{figure}[!ht]
	\begin{framed}
		\centering
		\begin{tabular}{p{.2\linewidth}p{.8\linewidth}}
			\tsf{SetTally:}& upon receiving ($result,\; \pi_{res}$) from administrator  \texttt{A}:  \\
			&   \texttt{Assert $admin$ = A}\\
			&   \texttt{Assert $T_{2}$ < $T$ < $T_{3}$}\\
			&   \texttt{Set $\vec{s}: = (result,\mg{\;Int\_Vsigns,}\; EncryptedVotes)$}\\
			&   \texttt{Assert zkSNARK.verify($\tsf{vk}_{C_{T}}$, $\vec{s}$, $\pi_{res}$)}\\
			&   \texttt{Set $tallyingResult := result$}
		\end{tabular}
	\end{framed}
	\caption{Pseudocode for \tsf{SetTally} function}
	\label{figTally}
\end{figure}

\subsubsection{Refunding.} Finally, at the end of tallying the result phase, the administrator and voters invoke \tsf{Refund} function to reclaim the collateral fee \mg {$F$ that} they deposited during deploying the smart contract and registering voters phases, respectively. Before  refunding the value, this function verifies that the time to set the tallying result has ended and the transaction sender indeed paid the deposit.

\section{Evaluation and \mg{Comparison}}\label{sec:evaluation}
We developed a prototype of our design to evaluate the gas cost and its scalability. The prototype is available as open-source on Github\footnote{\url{https://github.com/mhgharieb/zkSNARK-Open-Vote-Network}}. We utilize \ttt{circom} (v2.0) library\footnote{\url{https://docs.circom.io/}} to compile the arithmetic circuits \tsf{publicKeyGen}, \tsf{encryptedVoteGen}, and \tsf{Tallying}. During the protocol execution, we use \ttt{snarkjs} (v0.4.10) library\footnote{\url{https://github.com/iden3/snarkjs}} to handle the MPC-based setup ceremony for generating the proving and verifying keys. Also, we use it to generate the zk-SNARK proofs. We utilize \ttt{Tuffle} framework\footnote{\url{https://trufflesuite.com/}} (v5.4.24) to deploy and test the smart contracts.

The prototype includes three smart contracts: verifierMerkleTreeCon, verifierZKSNARKCon, and eVoteCon. The first two contracts are deployed as generic functions to verify the Merkle tree proof of membership and the zk-SNARKs proofs.
Table \ref{tab.cost} summarizes the gas units consumed by different functions in the contracts for executing the protocol for 40 voters. 

\begin{table}[t]
	\centering
	\caption{\small{The gas cost for functions in the voting contract}}
	\label{tab.cost}
	\begin{tabular}{cc}
		\toprule
		Function &Gas units\\
		\midrule
		verifierMerkleTreeCon &192,013\\
		verifierZKSNARKCon &1,346,155\\
		eVoteCon &1,474,818\\
		\hline
		\ttt{setVerifyingKey}($\tsf{vk}_{C_{PK}}$)&462,309\\
		\ttt{setVerifyingKey}($\tsf{vk}_{C_{V}}$)&2,253,285\\
		\ttt{setVerifyingKey}($\tsf{vk}_{C_{T}}$)&2,209,576\\
		\ttt{Register} &349,517\\
		\ttt{CastVote} &937,299\\
		\ttt{SetTally} &901,532\\
		\ttt{Refund} &52,253\\
		\bottomrule    
	\end{tabular}
\end{table}

\subsubsection{Scalability Experiments.} 
\begin{figure}[t]
	\includegraphics[width=1\textwidth]{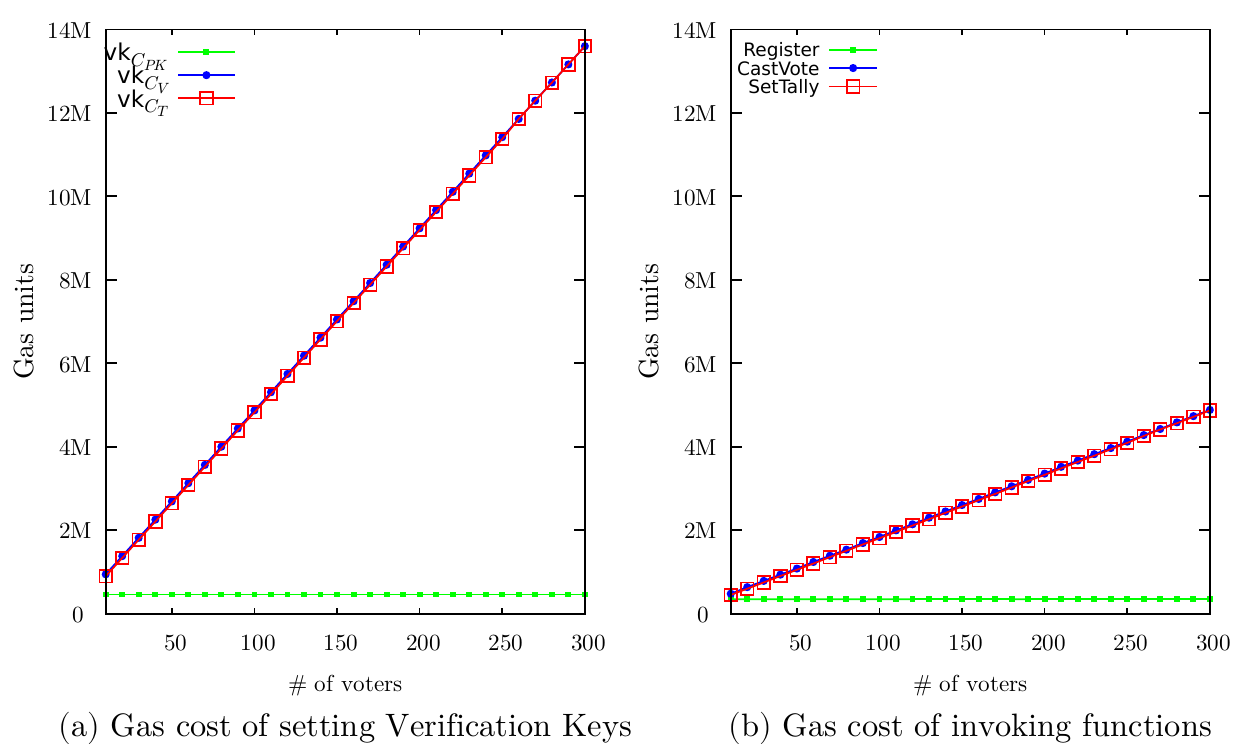}
	\caption{Gas cost in scalability experiments}
	\label{GasCost}
\end{figure}

We also conduct some experiments to estimate the maximum number of voters who can participate in the same election before exceeding the gas limit of the Ethereum block. To this end, we run the prototype with incremental steps starting from 10 voters up to \mg{300} voters. The gas cost of these experiments is shown in Fig. \ref{GasCost}. As depicted in Fig. \ref{GasCost}.a, setting the verifying keys $\tsf{vk}_{C_{V}}$ and $\tsf{vk}_{C_{T}}$ consumed an amount of gas units linearly with the number of voters. In contrast, the gas cost for setting the verifying key $\tsf{vk}_{C_{PK}}$ is approximately constant. This behavior is expected since the statement size $|\vec{s}|$ of the circuits \tsf{publicKeyGen} and \tsf{Tallying} is linearly proportional  to the number of voters. In contrast, it is constant in the case of \tsf{encryptedVoteGen} circuit. The same behavior is observed for invoking the smart contract functions as shown in Fig. \ref{GasCost}.b.

It is obvious that setting a verifying key can be made in multiple blocks. In contrast, running the smart contract function must be completed in the same block. Therefore, the gas cost of invoking the functions is the bottleneck against the scalability. With the current gas limit (30M gas units\footnote{\url{https://ethstats.net/}} on Jan. 16, 2022), this prototype can be scaled to around 2000 voters. 

\paragraph{Comparison with McCorry \etal \cite{mccorry}.} The design in \cite{mccorry} and ours provide the same properties specially the dispute-free property. In our design, we delegate the heavy computation to off-chain parties in  a verifiable manner. In contrast, in \cite{mccorry} all the computations are performed on-chain. For the sake of fairness, we compare the gas consumption by each voter and the administrator during a 40-voter election in the two designs as summarized in Table \ref{tab.compare}. In our design, the gas cost per voter and the administrator are around $40\%$ and $70\%$ of the other design, respectively. However, the gas cost increases rabidly with the number of voters in McCorry \etal design as deduced from Fig. 4 in \cite{mccorry}. Therefore, our design is more scalable.   

\begin{table}[t]
	\caption{Gas cost comparison between our implementation and \cite{mccorry}}
	\begin{center}
		\begin{tabular}{ccc}
			\toprule
			Sender&  Ours & \cite{mccorry}\\
			\midrule
			Voter~~~~&~~~~1,339,069~~~~~&~~~~3,323,642\\
			Admin~~~~&~~~8,719,141~~~&~~~~12,436,190\\
			\bottomrule    
		\end{tabular}
	\end{center}
	\label{tab.compare}
\end{table}

\paragraph{Comparison with Seifelnasr \etal \cite{seifelnasr}.} The implementation in \cite{seifelnasr}  delegates the tallying computation to an off-chain untrusted administrator along with a dispute phase to catch the misbehavior of a malicious administrator. In contrast, we delegate all the computation to off-chain parties without loosing the dispute-free property. Regarding the gas cost, after carefully reviewing the Github code of \cite{seifelnasr}, we found that their estimation does not take into account the `\ttt{verifyY}' step presented in the pseudocode of cast vote function in which the smart contract should verify that a voter \ttt{B} computes her blinding key $Y_B$ correctly. We argue that this step is the major factor on the gas cost of this function.
Moreover, they claimed that their design could be scaled theoretically to support $2^{256}$ voters because all transactions have constant gas cost except the two functions of registering voters and dispute phase which scales logarithmically with the number of voters since these functions verify the Merkle proof of membership. However, this claim also was based on ignoring the gas cost of the step `\ttt{verifyY}' which scales linearly with the number of voters.

\section{\mg{Further Scalability Improvement}}\label{furtherImprovement} 
Recall that the gas consumption \mg{by the functions: \ttt{setVerifyingKey}($\tsf{vk}_{C_{V}}$), \ttt{setVerifyingKey}($\tsf{vk}_{C_{T}}$), \tsf{CastVote}, and
		\tsf{SetTally}}, in the current design, increases linearly with the number of voters because the statement size of the zk-SNARK increases linearly with the number of voters. \mg{In this section, we present some modifications to} the circuits so that the size of the statement becomes fixed\mg{, hence \ttt{setVerifyingKey}($\tsf{vk}_{C_{V}}$) and \ttt{setVerifyingKey}($\tsf{vk}_{C_{T}}$) consume fixed gas units.  Regarding the gas consumption associated with  invoking the other smart contract functions, we propose some modifications on \tsf{Register}, \tsf{CastVote}, and \tsf{SetTally} functions so that they also consume fixed gas units independent of the number of voters.  As a result, the total gas paid by the administrator and each voter is constant, hence achieving very good scalability.}

\subsection{Fixed Statement Size}\label{fixedStatementSize}
Let \mg{$commit_{PK} = H(pk^y_0||pk^y_1||\cdots||pk^y_{n-1})$} denote the hash of the concatenation of \mg{the $y$-coordinates of the voters'} public keys. Recall that in \tsf{encryptedVoteGen} circuit, the statement $\vec{s}=$($V_B$, $i_B$, \{${pk}_i^{y}$\}) and the witness $\vec{w}=$($v_B$, $x_{B}$, \{${pk}_i^{x}$\}). By moving $\{{pk}_i^{y}\}$ from $\vec{s}$ to $\vec{w}$ and adding $commit_{PK}$ to $\vec{s}$, we can construct a new circuit \tsf{newEncryptedVoteGen} with new ($\vec{s'},\vec{w'}$) s.t.

\begin{align*}
	\vec{s'}=(V_B, commit_{PK}, i_B), &&	\vec{w'}=(v_B, x_{B}, \{{pk}_i^{x}\}, \{{pk}_i^{y}\})
\end{align*}

The \tsf{newEncryptedVoteGen} circuit is \tsf{encryptedVoteGen} circuit in addition to a new constraint $$\ttt{Assert}\; commit_{PK} =  H(pk^y_0||pk^y_1||\cdots||pk^y_{n-1})$$

Therefore, \tsf{newEncryptedVoteGen} has a fixed-size statement and independent of the number of voters, hence setting its verification key will consume fixed gas units. consequently, verifying the zk-SNARK proof will consume fixed gas units. In its turn, the smart contract should check the \mg{correctness of $commit_{PK}$} in \tsf{CastVote} function, since the public keys $PK_i$ are known. 

\mg{
Similarly, let $commit_{V} = H(V^y_0||V^y_1||\cdots||V^y_{n-1}||D_0||\cdots||D_l)$ denote the hash of the concatenation of the $y$-coordinates of the encrypted votes in addition to the integer numbers represented the sign-bits of the $x$-coordinates. Therefore, we can construct a new \tsf{Tallying} circuit with new ($\vec{s'},\vec{w'}$) s.t. $\vec{s'} = (res,commit_{V})$ and $\vec{w'} = ($\{$V_i^x$\},\{$V_i^y$\}). In its turn, the smart contract should check the correctness of $commit_{V}$ in \tsf{SetTallying} function since the encrypted votes $V_i$ are known.
}

\mg{
As a result, the total gas units consumed by \ttt{setVerifyingKey}($\tsf{vk}_{C_{V}}$) and \ttt{setVerifyingKey}($\tsf{vk}_{C_{T}}$) become constant. In contrast, the gas consumption by \tsf{CastVote} and \tsf{SetTallying} functions still increase linearly with the number of voters and depend on how much the used hash function cost. Accordingly, this approach is efficient and can support a high number of voters only if the used hash function is cheap on the Ethereum, \eg SHA-256.
}
\subsection{Fixed Gas Cost}\label{fixedGasCost}
\mg{Due to the structure of $commit_{PK}$ and $commit_{V}$,  they must be calculated after registering all voters and casting all encrypted votes, respectively. Therefore, the voter who casts the first encrypted vote has to pay the linearly increasing gas cost of calculating $commit_{PK}$ on behalf of all the other voters. Similarly, the administrator has to pay the linearly increasing gas cost associated with calculating $commit_{V}$. To achieve a constant gas paid by the administrator and each voter, we restructure $commit_{PK}$ and $commit_{V}$ so that they can be calculated in a progressive manner. Hence their gas cost is distributed among all voters.}
\mg{
Let the new $commit_{PK}$ be defined as follows: $$commit_{PK} = H(H(\cdots H(H(0||pk^y_0)||pk^y_1)||\cdots)||pk^y_{n-1})$$ Accordingly, the smart contract initializes $commit_{PK}:=0$ during its deployment, then during registering the public key $PK_i$, \tsf{Register} function updates  $commit_{PK} := H(commit_{PK}||pk^y_i)$. At the end of the voters registration phase, the new $commit_{PK}$ value is ready to be used by \tsf{CastVote} function to verify the correctness of the encrypted vote sent by each voter.
}

\mg{
Similarly, let the new $commit_{V}$ be defined as follows: $$commit_{V} = H(H(\cdots H(H(0||V^x_0||V^y_0)||V^x_1||V^y_1)||\cdots)||V^x_{n-1}||V^y_{n-1})$$ The smart contract initializes $commit_{V}:=0$, then during casting the encrypted vote $V_i$, \tsf{CastVote} function updates $commit_{V} := H(commit_{V}||V^x_i||V^y_i)$. By using both the two coordinates of $V_i$, there is no need of tracking the sign of the $x$-coordinates. At the end of the casting phase, the new $commit_{V}$ value is ready to be used by \tsf{SetTallying} function to verify the correctness of the tallying result sent by the administrator. 
}
\subsection{Performance Measurements}
\begin{figure}[t]
	\includegraphics[width=1\textwidth]{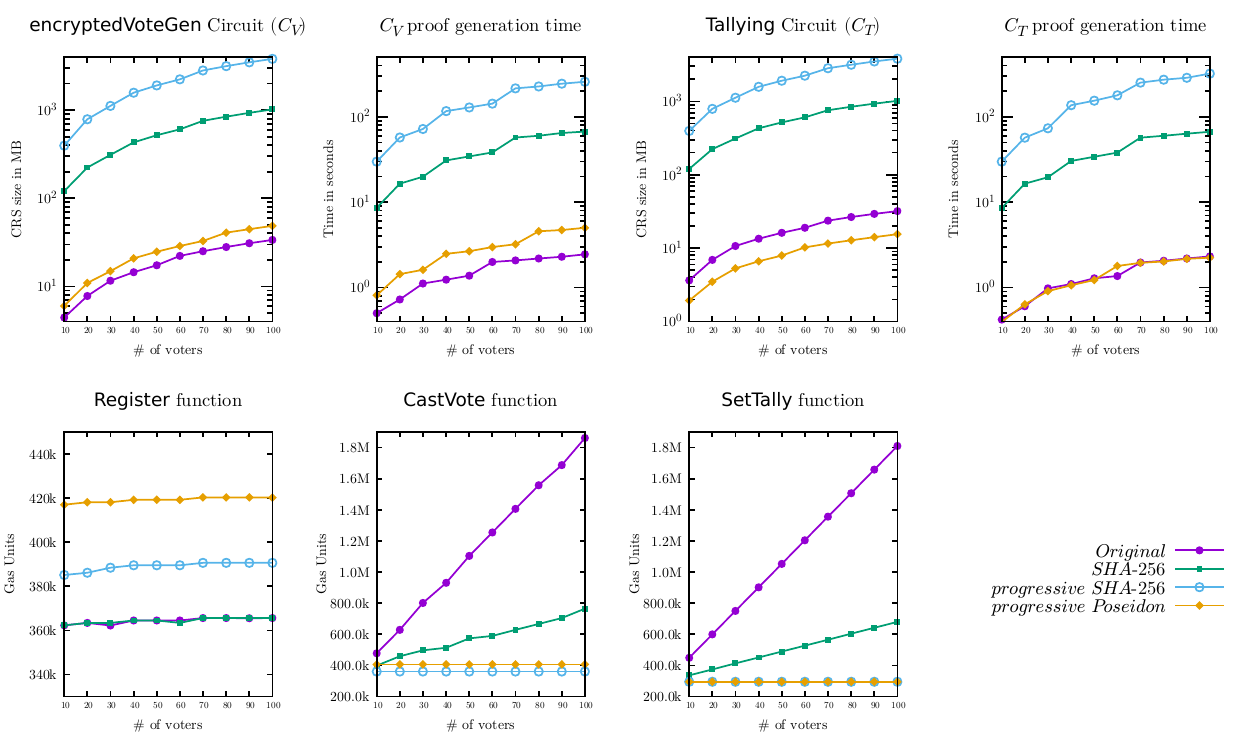}
	\caption{Performance evaluation of our modified designs. `\textit{Original}', `\textit{SHA-256}', `\textit{progressive SHA-256}', and `\textit{progressive Poseidon}' refer to our original design, the modified designs in Sec. \ref{fixedStatementSize} using SHA-256, and Sec. \ref{fixedGasCost} using the hash functions SHA-256, and Poseidon, respectively. }
	\label{enhanced}
\end{figure}

\mg{
We evaluate the performance of the two modified designs in terms of the size of the common reference string (CRS)  for \tsf{encryptedVoteGen} and \tsf{SetTallying} circuits (\ie the proving and verification keys sizes); their average proof generation time; and the gas consumption. associated with invoking the three functions: \tsf{Register}, \tsf{CastVote}, and \tsf{SetTally}.
In particular, we evaluate our implementation using SHA-256 in Sec. \ref{fixedStatementSize} (referred as \textit{SHA-256}), and SHA-256 and Poseidon \cite{poseidon}  in Sec. \ref{fixedGasCost} (referred as \textit{progressive SHA-256} and \textit{progressive Poseidon}).}
\mg{As depicted in Fig. \ref{enhanced}, the CRS size and the average proof generation time for both circuits increase linearly with the number of voters in our original and modified designs. The highest value corresponds to the \textit{progressive SHA-256} hashing case. As expected,} SHA-256 cannot be presented by a friendly arithmetic circuit, \ie it generates a high number of constraints, hence a large proving \mg{key size} and high  zk-SNARK proof generation time. \mg{In contrast, Poseidon hash function is an arithmetic circuit friendly hash function.}
\mg{Regarding the gas cost, \tsf{Register} function consumes  fixed gas units in all designs independent of the number of voters. However, it is the highest in \textit{progressive Poseidon} hashing case. For both \tsf{CastVote}, and \tsf{SetTally} functions, their gas costs for \textit{SHA-256} increase linearly with the number of voters, but \textit{SHA-256} is still more scalable than our original design. In contrast, the two functions consume a fixed gas units in \textit{progressive SHA-256} hashing and \textit{progressive Poseidon} hashing as desired. However, the gas cost of \textit{progressive Poseidon} hashing is slightly higher for \tsf{CastVote} function.
The decision of which design to be used should be taken by the administrator and the voters since it presents a trade-off between the proof generation time and the money spent in the form of the gas fees.
}

\section{Conclusion}\label{sec:conclusion}
We presented a dispute-free implementation for the Open Vote Network protocol as smart contracts in which all the heavy computations are performed off-chain. Then, we utilized zk-SNARKs to verify that all off-chain computations are performed correctly by their corresponding parties. Moreover, we developed a prototype of our design to assess its performance. \mg{Then,  we enhanced its scalability by utilizing progressive hashing to achieve fixed zk-SNARK statement sizes and distribute the gas costs among all voters. As a result, the total gas paid by the administrator and each voter is constant, hence achieving very good scalability.} 


\bibliography{bib/bibliography}

\begin{thebibliography}{10}
\providecommand{\url}[1]{\texttt{#1}}
\providecommand{\urlprefix}{URL }
\providecommand{\doi}[1]{https://doi.org/#1}

\bibitem{adida2008helios}
Adida, B.: {Helios: Web-based Open-Audit Voting}. In: USENIX security
  symposium. vol.~17, pp. 335--348 (2008)

\bibitem{math9233022}
Bellés-Muñoz, M., Whitehat, B., Baylina, J., Daza, V., Muñoz-Tapia, J.L.:
  {Twisted Edwards Elliptic Curves for Zero-Knowledge Circuits}. Mathematics
  \textbf{9}(23) (2021)

\bibitem{eip-197}
Buterin, V., Reitwiessner, C.: {EIP}-197: {Precompiled} contracts for optimal
  ate pairing check on the elliptic curve alt\_bn128 (2017),
  \url{https://eips.ethereum.org/EIPS/eip-197}

\bibitem{galal2019efficient}
Galal, H.S., ElSheikh, M., Youssef, A.M.: {An Efficient Micropayment Channel on
  Ethereum}. In: Data Privacy Management, Cryptocurrencies and Blockchain
  Technology -- CBT 2019. LNCS, vol. 11737, pp. 211--218. Springer (2019)

\bibitem{poseidon}
Grassi, L., Khovratovich, D., Rechberger, C., Roy, A., Schofnegger, M.:
  {Poseidon: A New Hash Function for {Zero-Knowledge} Proof Systems}. In: 30th
  USENIX Security Symposium (USENIX Security 21). pp. 519--535 (2021)

\bibitem{GRITZALIS2002539}
Gritzalis, D.A.: Principles and requirements for a secure e-voting system.
  Computers \& Security  \textbf{21}(6),  539--556 (2002)

\bibitem{groth2004efficient}
Groth, J.: {Efficient Maximal Privacy in Boardroom Voting and Anonymous
  Broadcast}. In: International Conference on Financial Cryptography -- FC
  2004. LNCS, vol.~3110, pp. 90--104. Springer (2004)

\bibitem{groth2010short}
Groth, J.: {Short Pairing-Based Non-interactive Zero-Knowledge Arguments}. In:
  Advances in Cryptology - ASIACRYPT 2010. LNCS, vol.~6477, pp. 321--340.
  Springer (2010)

\bibitem{groth16}
Groth, J.: {On the Size of Pairing-Based Non-interactive Arguments}. In:
  Advances in Cryptology – EUROCRYPT 2016. LNCS, vol.~9666, pp. 305--326.
  Springer (2016)

\bibitem{hao2010anonymous}
Hao, F., Ryan, P.Y., Zieli{\'n}ski, P.: Anonymous voting by two-round public
  discussion. IET Information Security  \textbf{4}(2),  62--67 (2010)

\bibitem{jivsov_compact_2014}
Jivsov, A.: {Compact representation of an elliptic curve point} (Mar 2014),
  \url{https://tools.ietf.org/id/draft-jivsov-ecc-compact-05.html}

\bibitem{10.1007/3-540-45664-3_10}
Kiayias, A., Yung, M.: {Self-tallying Elections and Perfect Ballot Secrecy}.
  In: Naccache, D., Paillier, P. (eds.) {Public Key Cryptography -- PKC 2002}.
  LNCS, vol.~2274, pp. 141--158. Springer (2002)

\bibitem{8405627}
Kshetri, N., Voas, J.: {Blockchain-Enabled E-Voting}. IEEE Software
  \textbf{35}(4),  95--99 (2018)

\bibitem{9149825}
Li, H., Li, Y., Yu, Y., Wang, B., Chen, K.: {A Blockchain-Based Traceable
  Self-Tallying E-Voting Protocol in AI Era}. IEEE Transactions on Network
  Science and Engineering  \textbf{8}(2),  1019--1032 (2021)

\bibitem{9031381}
Li, Y., Susilo, W., Yang, G., Yu, Y., Liu, D., Du, X., Guizani, M.: {A
  Blockchain-Based Self-Tallying Voting Protocol in Decentralized IoT}. IEEE
  Transactions on Dependable and Secure Computing  \textbf{19}(1),  119--130
  (2022)

\bibitem{Maaten2004}
Maaten, E.: Towards remote e-voting: Estonian case. In: Prosser, A., Krimmer,
  R. (eds.) Electronic voting in Europe - Technology, law, politics and
  society, workshop of the ESF TED programme together with GI and OCG. pp.
  83--90. Gesellschaft für Informatik e.V., Bonn (2004)

\bibitem{mccorry}
McCorry, P., Shahandashti, S.F., Hao, F.: {A Smart Contract for Boardroom
  Voting with Maximum Voter Privacy}. In: International Conference on Financial
  Cryptography and Data Security. LNCS, vol. 10322, pp. 357--375. Springer
  (2017)

\bibitem{murtaza2019blockchain}
Murtaza, M.H., Alizai, Z.A., Iqbal, Z.: {Blockchain Based Anonymous Voting
  System Using zkSNARKs}. In: 2019 International Conference on Applied and
  Engineering Mathematics (ICAEM). pp. 209--214. IEEE (2019)

\bibitem{parno2013pinocchio}
Parno, B., Howell, J., Gentry, C., Raykova, M.: {Pinocchio: Nearly practical
  verifiable computation}. In: 2013 IEEE Symposium on Security and Privacy. pp.
  238--252. IEEE (2013)

\bibitem{sasson2014zerocash}
Sasson, E.B., Chiesa, A., Garman, C., Green, M., Miers, I., Tromer, E., Virza,
  M.: {Zerocash: Decentralized Anonymous Payments from Bitcoin}. In: 2014 IEEE
  Symposium on Security and Privacy. pp. 459--474. IEEE (2014)

\bibitem{seifelnasr}
Seifelnasr, M., Galal, H.S., Youssef, A.M.: {Scalable Open-Vote Network on
  Ethereum}. In: International Conference on Financial Cryptography and Data
  Security. LNCS, vol. 12063, pp. 436--450. Springer (2020)

\end{thebibliography}


\clearpage

\end{document}